# Chemical Short-Range Order Controls Deformation Pathways in a Complex Concentrated Alloy

Angelo F. Andreoli[a,b,c,*], Guilherme C. Stumpf[d,e], Gabriela B. Ribeiro[d], Maria F. L. Valverde[a], Gustavo Bertoli[a], Vinícius P. Bacurau[d], David D. S. Silva[a], Pedro H. F. Oliveira[f], Eric M. Mazzer[a,d], Mamta Silwal[g], Garritt J. Tucker[g,h], Rodrigo Freitas[e], Martin Sahlberg[b,c], Daniel Miracle[i], Francisco G. Coury[a,d]

[a] Department of Materials Engineering, Federal University of São Carlos, São Carlos, SP, Brazil
[b] Department of Chemistry, Uppsala University, Uppsala, Sweden
[c] Wallenberg Initiative Materials Science for Sustainability, Uppsala University, Uppsala, Sweden
[d] Graduate Program in Materials Science and Engineering, Federal University of São Carlos, São Carlos, SP, Brazil
[e] Department of Materials Science and Engineering, Massachusetts Institute of Technology, Cambridge, MA, USA
[f] Department of Materials, The University of Manchester, Manchester, United Kingdom
[g] Department of Physics and Astronomy, Baylor University, Waco, TX, USA
[h] Department of Mechanical Engineering, Baylor University, Waco, TX, USA
[i] AF Research Laboratory, Materials and Manufacturing Directorate, Wright-Patterson AFB, OH, USA

* Corresponding author: A. F. Andreoli (angelo.fernandes-andreoli@kemi.uu.se)

**Abstract** – Chemical short-range order (CSRO) is an intrinsic feature of complex concentrated alloys, yet its influence on deformation mechanisms is controversial because of the inconclusive state of concurrent CSRO quantification during deformation. Here, we demonstrate that CSRO controls deformation-induced phase transformation in a $Co_{30}Cr_{40}Ni_{30}$ alloy. Recrystallized and aged conditions with comparable grain sizes, texture, and phase constitution were used to isolate the effect of atomic-scale chemical ordering. Calorimetry reveals reversible signatures of chemical ordering and disordering associated with the two thermal states. Synchrotron X-ray diffraction and electron microscopy show that the CSRO-enriched condition develops substantially less deformation-induced *hcp* martensite than the CSRO-lean condition at both room and cryogenic temperatures, despite broadly similar tensile strength. Hybrid Monte Carlo/molecular dynamics simulations further show that the CSRO configuration increases the stable and unstable stacking-fault energies, thereby modifying the energetic landscape for Shockley partial-dislocation activity and *fcc→hcp* transformation. The combined results establish a mechanistic link between local chemical correlations, planar-fault energetics, and deformation-pathway selection. CSRO therefore provides an additional materials-design variable, beyond alloy composition, for controlling deformation mechanisms in complex concentrated alloys.

## Introduction

Stacking-fault energy (SFE) engineering and deformation-mechanism control is a central strategy for designing structural materials with high strength, ductility, and damage tolerance [1]. In general, SFE and deformation pathways can be tuned through composition [2], temperature [3], grain size [4], and crystallographic texture [5]. Complex concentrated alloys (CCAs) introduce an additional structural variable: the local chemical environment [6,7]. Even when constituent elements remain randomly distributed on average, preferential atomic interactions can produce chemical short-range order (CSRO) [8], generating local correlations that modify the energetic landscape of the solid solution [9]. CSRO has been observed over a broad temperature range in CoCrNi-based alloys and investigated using calorimetry [10], electrical resistivity [11], transmission electron microscopy (TEM) [12], atom probe tomography [13], neutron scattering [14], and atomistic simulations [15–19].

Unlike long-range order, CSRO does not produce a distinct periodic structure and is therefore inherently statistical. It is commonly quantified using the Warren–Cowley parameter (WC), which describes deviations in the local chemical environment from a random solid solution [20]. Although the thermodynamic origin and atomic-scale nature of CSRO in CCAs are increasingly understood [21,22], its consequences for plastic deformation remain unresolved [23]. Whether local chemical correlations can act as an independent variable for selecting competing deformation mechanisms has not been established experimentally. Reports of CSRO-dependent changes in strength and dislocation behavior remain inconsistent [24–26] while direct experimental evidence linking CSRO to deformation-induced transformation (TRIP) or twinning (TWIP) is scarce [27].

This question is particularly relevant to alloys in which multiple deformation mechanisms compete. In face-centered-cubic (*fcc*) metals with low SFE, perfect dislocations can dissociate into Shockley partials [28], and the resulting stacking faults can promote deformation twinning or successive fault accumulation leading to *fcc*→ hexagonal closed-packed (*hcp*) martensitic transformation [7,29]. The SFE therefore provides a critical thermodynamic parameter linking atomic-scale structure to macroscopic deformation pathways. In CoCrNi-based alloys, low SFE promotes TWIP and TRIP mechanisms [30,31], with TRIP becoming particularly pronounced at cryogenic temperatures [3], and the interplay between these mechanisms contributes to their exceptional damage tolerance [32–34]. Atomistic calculations further indicate that CSRO modifies stacking-fault energetics [35] and can increase the barriers associated with dislocation motion in CoCrNi [6,36]. Whether these changes are sufficient to alter the competition between deformation mechanisms in a bulk polycrystalline alloy, however, remains an open question.

Here, we address this question using a $Co_{30}Cr_{40}Ni_{30}$ alloy, selected for its low SFE and pronounced susceptibility to TRIP activity [37]. Two thermally processed states were designed to provide comparable grain structure, texture, and phase constitution while producing distinct CSRO states: a recrystallized and quenched condition with comparatively low CSRO and an aged condition enriched in CSRO. This approach allows the influence of atomic-scale chemical correlations to be separated from major microstructural variables. We combine calorimetry, tensile testing at ambient and cryogenic temperatures, synchrotron X-ray diffraction (XRD), electron backscatter diffraction (EBSD), and TEM to establish the effect of CSRO on deformation pathways. Hybrid Monte Carlo/molecular dynamics (MC/MD) simulations and generalized planar-fault-energy calculations are then used to connect the experimentally observed deformation behavior with the underlying atomic-scale energetics.

We show that CSRO substantially suppresses deformation-induced *fcc→hcp* transformation at both 300 and 173 K, despite broadly similar initial microstructures and tensile strengths. The simulations reveal that the CSRO configuration increases both stable and unstable stacking-fault energies, modifying the energetic accessibility of Shockley partial-dislocation activity and thereby reducing the propensity for the sequential faulting associated with *fcc→hcp* transformation. These results establish a mechanistic connection between local chemical correlations, planar-fault energetics, and deformation-pathway selection, demonstrating that CSRO provides an additional degree of freedom for controlling deformation mechanisms in CCAs.

## Results

### *Microstructure of recrystallized and aged conditions*

**Figure 1** presents the microstructure and phase constitution of the recrystallized and aged conditions. The EBSD inverse pole figure (IPF) maps (**Figs. 1a and 1b**) reveal fully recrystallized equiaxed microstructures with numerous annealing twins in both conditions. Excluding twin boundaries, the mean grain sizes are 50.9 ± 22.7 μm and 45.5 ± 17.0 μm (± standard deviation) for the recrystallized and aged conditions, respectively (**Figs. 1e and 1f**). Although the recrystallized condition exhibits slightly larger grain sizes and broader distributions, these differences fall within the corresponding standard deviations and are not expected to influence either the tensile response or TRIP behavior in CoCrNi-based alloys [4]. Consequently, the two microstructures are considered equivalent for the present analysis. Furthermore, the misorientation angle distributions (**Figs. 1h and 1i**) closely follow the Mackenzie distribution, indicating essentially random textures in both conditions.

The EBSD phase maps (**Figs. 1c–d**) indicate a single-phase *fcc* microstructure for both conditions, in agreement with the XRD patterns shown in **Fig. 1g**. No diffraction peaks indicative of secondary phases was detected. Additional texture analysis, microstructural characterization, EDS maps, and composition measurements are provided in **Figs. S2–S5** and **Table S1** of the Supplementary Information.

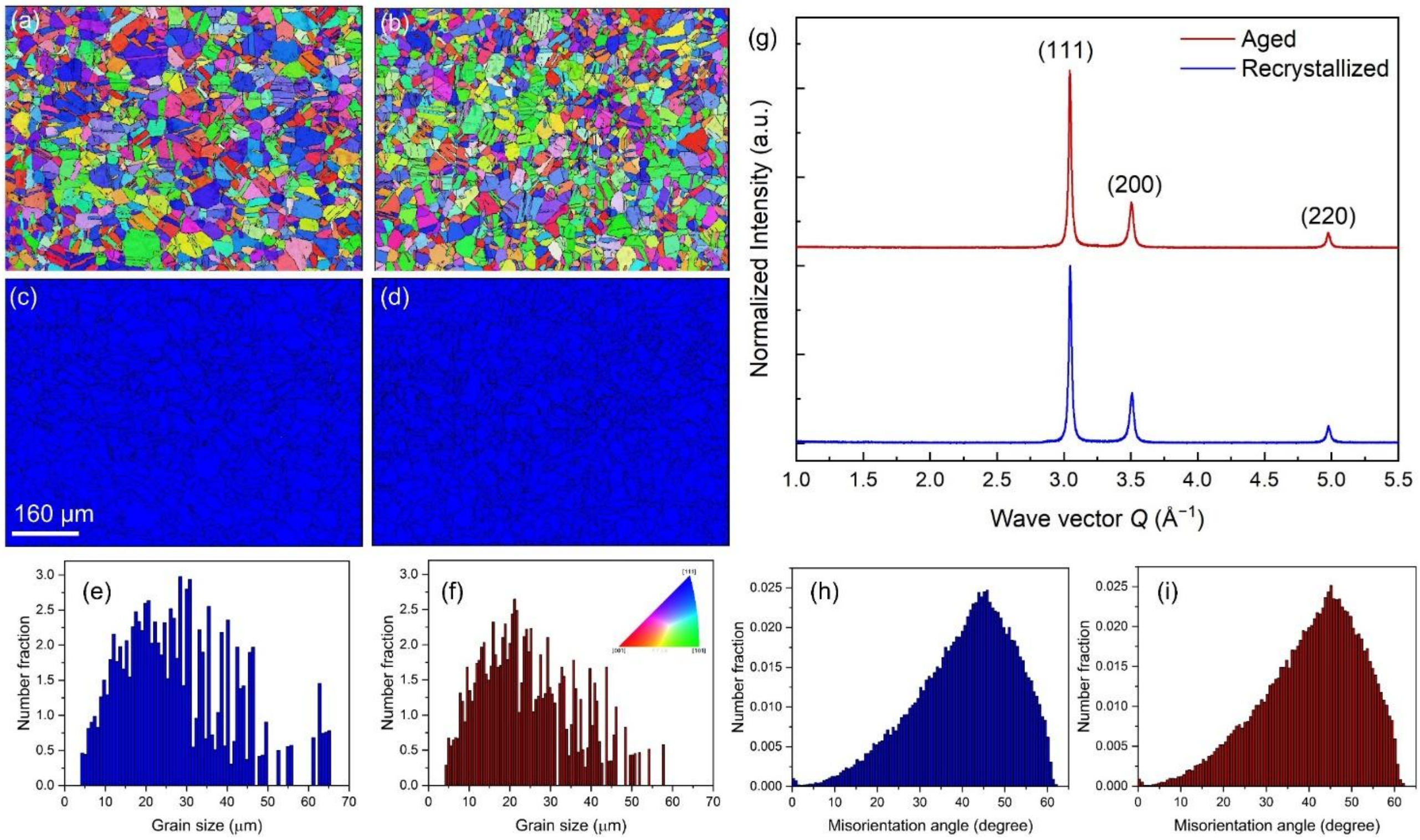

**Figure 1**. **The microstructure and phase constitution of the $Co_{30}Cr_{40}Ni_{30}$ alloy with distinct thermal histories**. (**a, b**) IPFs of recrystallized and aged conditions, respectively; (**c, d**) phase maps of recrystallized and aged conditions, respectively; (**e, f**) grain size distribution plots of recrystallized and aged conditions, respectively; (**g**) XRD patterns of recrystallized and aged conditions; (**h, i**) grain misorientation angle distribution of recrystallized and aged conditions, respectively.

***Correlation between heat capacity and CSRO***

Based on the specific heat capacity ($c_p$) curves obtained for a recrystallized sample (**Fig. 2a**), in the first heating cycle, an anomalous decrease in heat capacity was observed beginning slightly below 773 K and reaching a minimum around 823 K, indicating an exothermic heat flux process in the differential scanning calorimeter (DSC). In contrast, the aged sample showed a pronounced increase in $c_p$ around 823 K (**Fig. 2b**).

This behavior is consistent with previous calorimetric investigations of CSRO in CoCrNi-based alloys [10,22]. In the recrystallized condition (CSRO-lean), the exothermic anomaly observed during the first heating cycle (**Fig. 2a**) arises from the formation of energetically favorable local atomic correlations, once diffusion becomes appreciable, as the initially disordered alloy evolves toward a lower-enthalpy ordered state [21]. In contrast, the aged condition exhibits an endothermic anomaly near 823 K, corresponding to the dissolution of pre-existing CSRO and the recovery of configurational entropy. These interpretations are supported by previous calorimetry and atomistic simulations, which demonstrate that this temperature range is associated with CSRO formation and dissolution in CoCrNi-based alloys [8,16].

The disappearance of an anomalous $c_\mathrm{p}$ decrease during the second heating cycle further confirms that they originate from reversible ordering phenomena rather than phase transformations [21]. After experiencing comparable ordering–disordering histories, both conditions display nearly identical heat capacity responses. This reversible calorimetric signature has been established as a robust indirect fingerprint of CSRO in CoCrNi-based alloys [10,22].

Integration of the calorimetric peaks yields enthalpy changes of −7.88 J $g^{-1}$ for CSRO formation in the recrystallized condition (**Fig. 2a**) and +3.07 J $g^{-1}$ for CSRO dissolution in the aged condition (**Fig. 2b**). These results confirm that the aging treatment produced a CSRO-enriched state, whereas the recrystallized condition remained comparatively CSRO-lean before deformation testing [8].

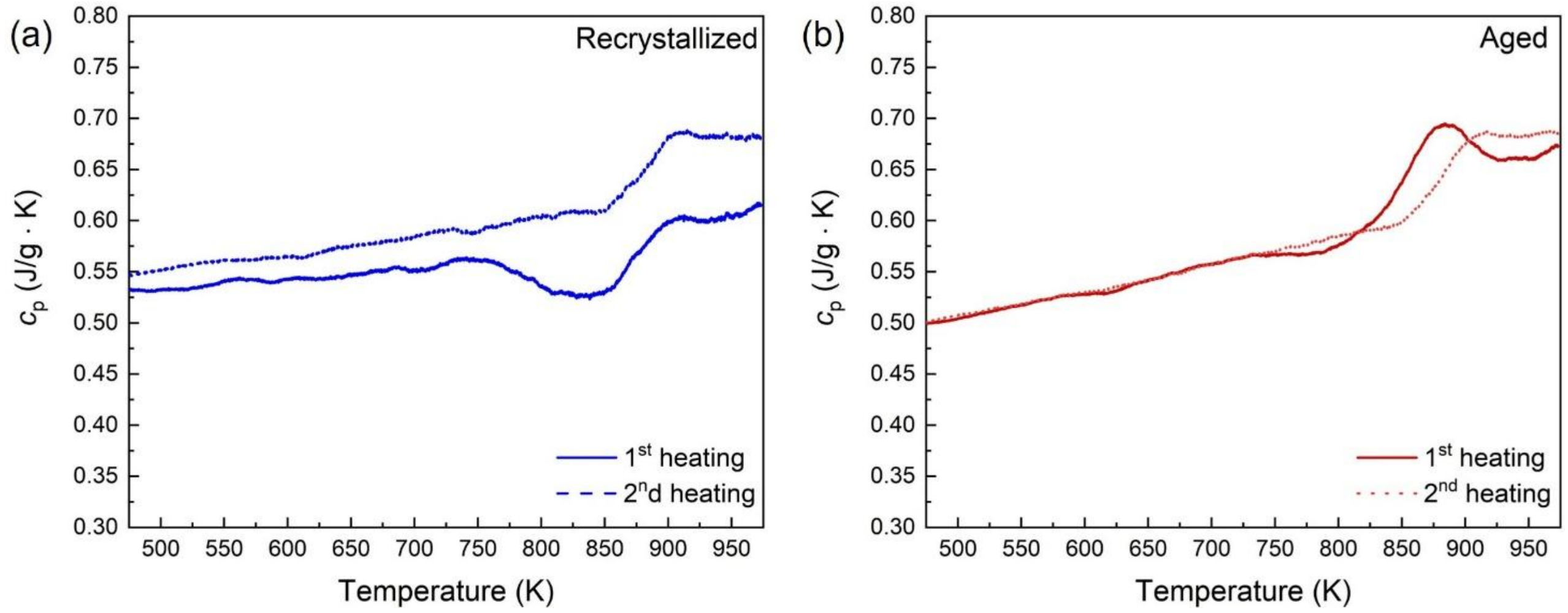

**Figure 2**. **The temperature-dependent specific heat capacity $c_\mathrm{p}(T)$ of the $Co_{30}Cr_{40}Ni_{30}$ alloy in the (a) recrystallized, and (b) aged conditions**. The first and second heating cycles are shown.

***Mechanical behavior of recrystallized and aged conditions***

The tensile behavior of the recrystallized (REC) and aged (CSRO) conditions evaluated at 300 K and 173 K are displayed in **Fig. 3**. At 300 K, both conditions show comparable yield strengths $\sigma_t$, with values of 338 ± 5 MPa for the REC300 and 342 ± 17 MPa for CSRO300. The ultimate tensile strength (UTS) for the recrystallized condition (average value 736 ± 15 MPa) is within the estimated error bar of the aged condition (726 ± 15 MPa). At the same time, both states display similar fracture strains, $\varepsilon_t$ of approximately 68 ± 5 %. These observations are consistent with the nearly overlapping engineering stress–strain curves shown in **Fig. 3a** at 300 K.

Reducing the test temperature to 173 K results in a clear increase in strength and elongation at fracture under both conditions. The yield strength increases to 391 ± 28 MPa for the REC173 and 393 ± 22 MPa for CSRO173, accompanied by a large rise in UTS to 869 ± 13 MPa and 841 ± 3 MPa, respectively (**Fig. 3c**). Remarkably, the fracture strain also increases at 173 K, reaching 84 ± 7 % for the REC173 and 74 ± 2 % for the CSRO173, in agreement with the extended uniform deformation seen in **Figs. 3a** and **3b**.

The true stress–true strain curves (**Fig. 3b**) show sustained strain hardening to large plastic strains in all cases, with the recrystallized condition assessed at 173 K (REC173) achieving the highest true stress before fracture. This enhanced work-hardening capability at low temperature is further reflected in the strain-hardening rate curves (**Fig. 3d**), which show higher $\delta\sigma_t/\delta\varepsilon_t$ values and a more extended plateau for the recrystallized condition compared with the aged condition, particularly at 173 K.

The tensile data demonstrate that lowering the deformation temperature increases both strength and ductility, while the recrystallized condition shows higher work-hardening capacity and fracture strain than the aged condition at low temperatures.

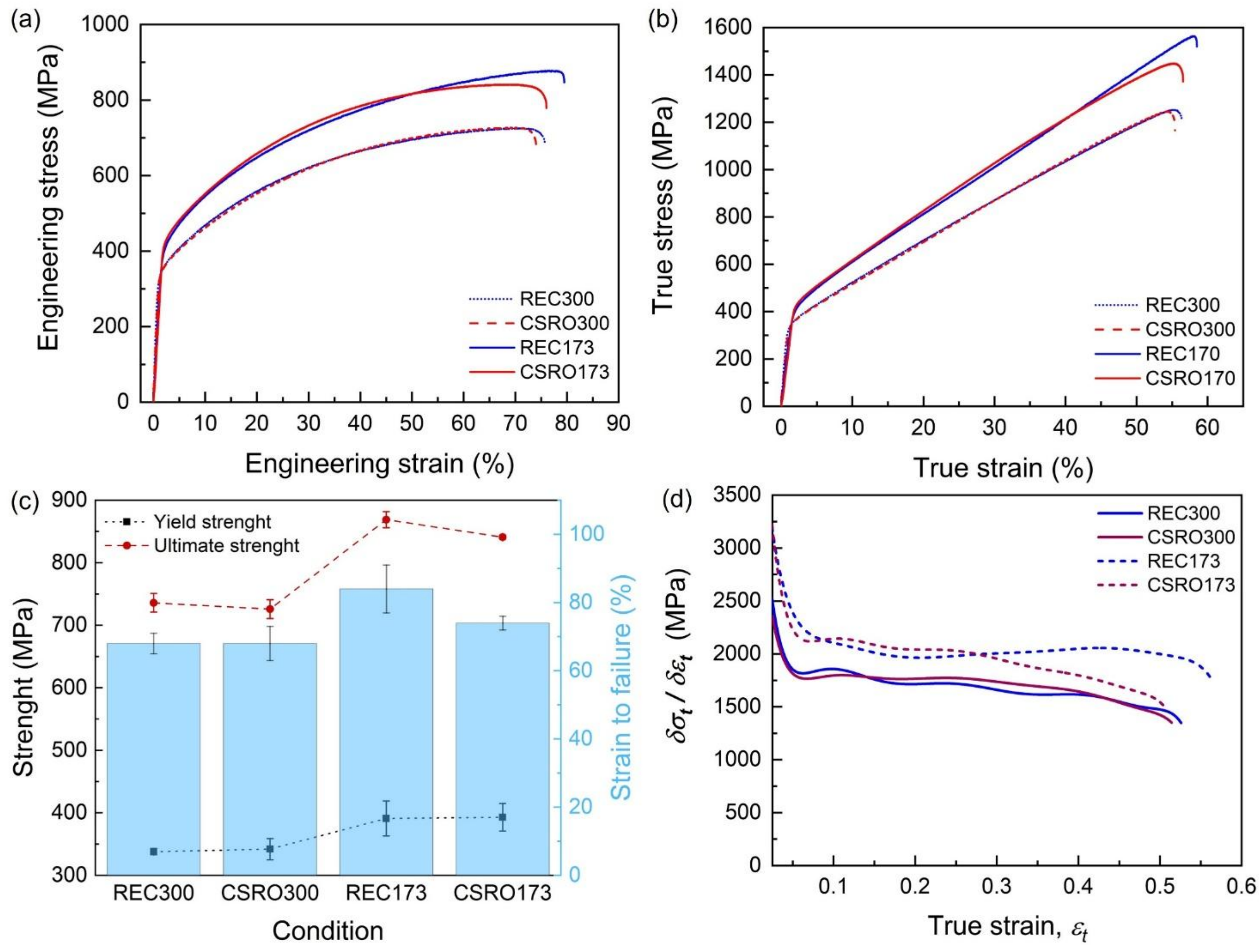

**Figure 3**. **Tensile behavior of the recrystallized (REC) and aged (CSRO) conditions tested at 300 K and 173 K**. (**a**) Engineering stress–strain curves; (**b**) True stress–true strain curves; (**c**) tensile properties of the recrystallized and aged conditions tested at 300 K and 173 K, including yield strength ($\sigma_t$), ultimate tensile strength (UTS), and fracture strain $\varepsilon_t$; (**d**) evolution of the instantaneous strain-hardening rate ($\delta\sigma_t/\delta\varepsilon_t$) as a function of true strain.

***Microstructure of deformed samples***

**Figure 4** presents the synchrotron (SXRD) patterns measured at room temperature and acquired after tensile deformation at 300 K and 173 K of recrystallized and aged conditions, together with the corresponding determined phase volume fractions. For comparison purposes, the diffractograms' intensities were normalized. No diffraction peaks other than the *fcc* and *hcp* phase are seen in any tested condition (**Figs. 4a-b** and **S6**). The diffraction profiles of specimens deformed at 300 K (**Fig. 4a**) are dominated by reflections indexed to the *fcc* phase, with additional weak peaks associated with the *hcp* phase. After deformation at 300 K, the recrystallized condition (REC300) shows discernible *hcp* reflections, while the aged condition (CSRO300) shows substantially reduced *hcp* peak intensity. Extended Q-range SXRD diffractograms are provided in **Fig. S6** in Supplementary Information.

For samples tested at 173 K (**Fig. 4b**), the intensity of the *hcp* reflections increases markedly for both microstructural conditions, showing an enhanced TRIP effect at lower testing temperature, as expected [38]. The recrystallized condition (REC173) displays the most pronounced *hcp* peaks, while the aged condition (CSRO173) again shows weaker *hcp* contributions under otherwise identical deformation conditions.

The phase volume fractions measured from the SXRD data are summarized in **Fig. 4c**. After tensile deformation at 300 K, the recrystallized condition (REC300) holds 5 ± 0.4 % *hcp* phase, while the aged condition (CSRO300) shows a significantly lower *hcp* fraction of 1 ± 0.2 %. At 173 K, the *hcp* phase fraction increases to 11 ± 0.2 % in the recrystallized condition (REC173) and to 6 ± 0.2 % in the aged condition (CSRO173). In all conditions, the balance of the microstructure consists of the *fcc* phase. Overall, **Fig. 4** indicates that the extent of deformation-induced *hcp* transformation increases by decreasing testing temperature and is consistently lower in the aged condition compared with the recrystallized state at both test temperatures. The volume fractions are summarized in **Table 1**. For comparison to the SXRD data, in-house measured XRD is displayed in **Fig. S7**, in the Supplementary Information.

**Table 1. The phase constitution of the $Co_{30}Cr_{40}Ni_{30}$ alloy after tensile testing at 300 and 173 K determined from EBSD and SXRD techniques.**

| | Phase fraction (%) | REC300 | CSRO300 | REC173 | CSRO173 |
|---|---|---|---|---|---|
| **EBSD** | *fcc* | 91 | 99 | 80 | 94 |
| | *hcp* | 9 | 1 | 20 | 6 |
| **SXRD** | *fcc* | 95 | 99 | 89 | 94 |
| | *hcp* | 5 | 1 | 11 | 6 |

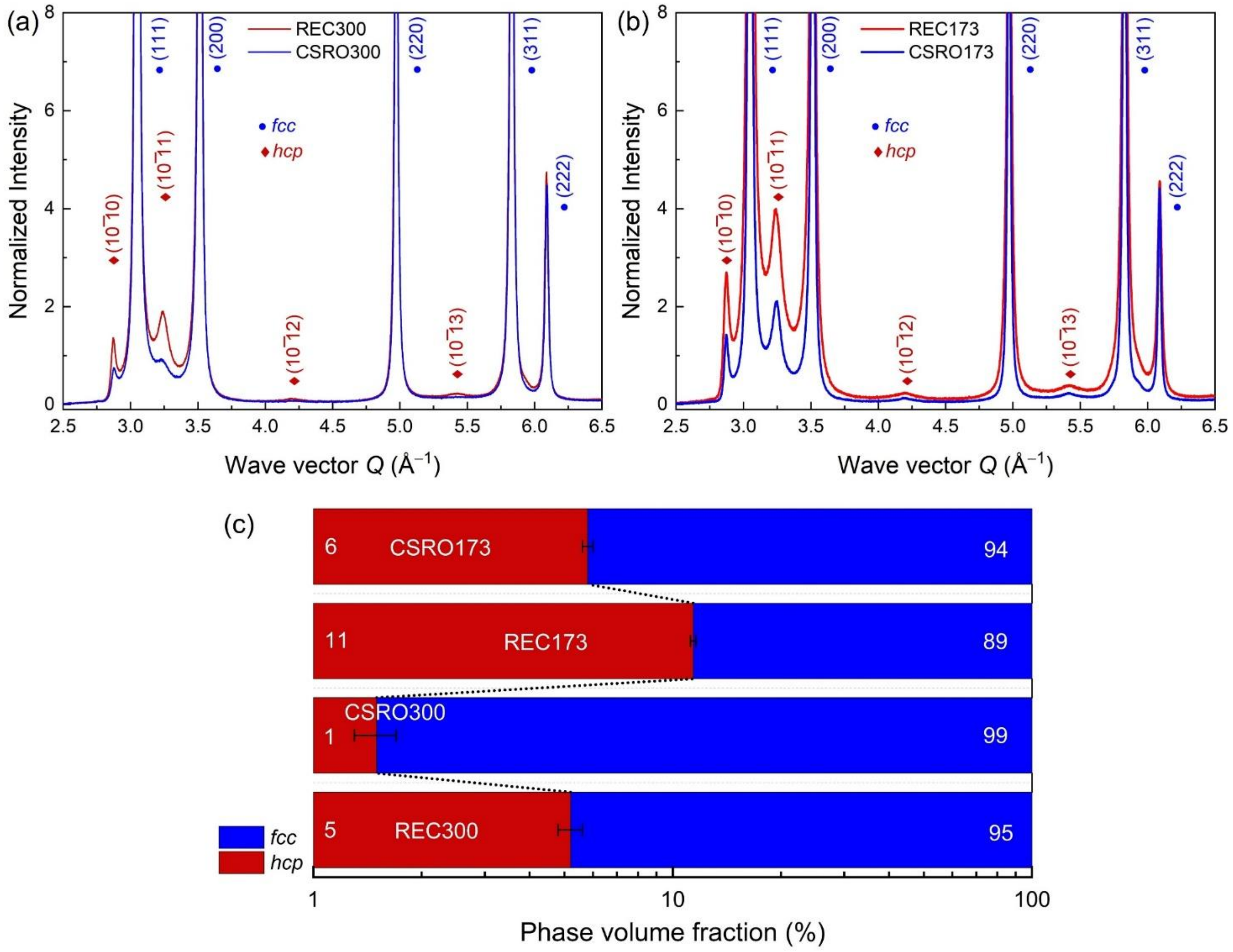

**Figure 4. Synchrotron X-ray diffraction patterns obtained after tensile deformation of the recrystallized and aged conditions and determined phase volume fraction.** (**a**) SXRD patterns measured at 300 K; (**b**) SXRD patterns measured at 173 K. Reflections corresponding to the *fcc* and *hcp* phases are indicated; (**c**) phase volume fractions of the *fcc* and *hcp* phases determined from the diffraction data for each condition and testing temperature. The x-axis is presented on a logarithmic scale to better highlight the volume fraction of the deformation-induced *hcp* product phase.

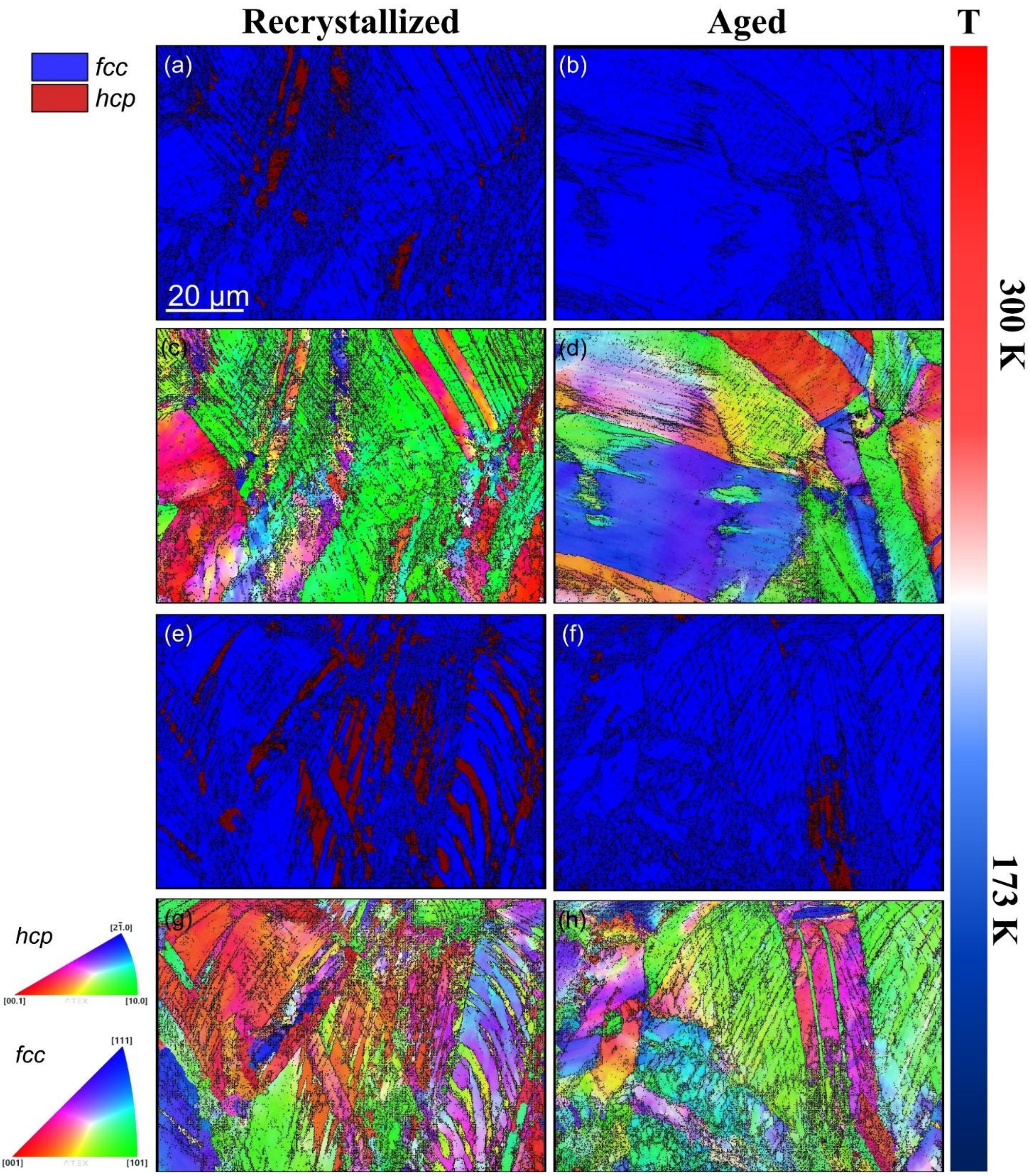

**Figure 5**. **EBSD maps of post-deformation $Co_{30}Cr_{40}Ni_{30}$ samples**. (**a**, **b**) phase distribution maps of recrystallized and aged conditions tested at 300 K, respectively; (**c**, **d**) (100) IPF maps of recrystallized and aged conditions tested at 300 K, respectively; (**e**, **f**) phase distribution maps of recrystallized and aged conditions tested at 173 K, respectively; (**g**, **h**) (100) IPF maps of recrystallized and aged conditions tested at 173 K, respectively.

**Figure 5** shows the EBSD phase and IPF maps obtained after deformation under different conditions. In the recrystallized condition evaluated at 300 K (**Figs. 5a and 5c**), a heterogeneous deformation microstructure is seen, characterized by numerous intragranular lamellae and deformation bands. EBSD phase indexing shows a measurable fraction *of* hcp lamellae embedded within the *fcc* matrix, consistent with TRIP activity. Quantitative analysis reveals that 9 % *hcp* phase is formed during deformation at 300 K (**Table 1**), and the lamellae follow well-defined {111} *fcc* habit planes, consistent with the *fcc→hcp* transformation mechanism.

The aged condition, evaluated at room temperature (**Figs. 5b and 5d**), however, shows a distinctly reduced volume fraction of the *hcp* phase. Although the deformation bands and planar slip traces are still visible, the *hcp* fraction drops to ~1% according to EBSD indexing (**Table 1**). This suppression of transformation activity provides evidence that the ageing treatment, and the associated development of CSRO, modifies the deformation pathway. The qualitative agreement trend between EBSD and SXRD phase quantification (**Fig. 4c**) confirms that this reduction is not an artifact of indexing resolution or Kikuchi band quality.

Temperature further amplifies these trends. When tested at 173 K, the recrystallized condition shows a substantial increase in strain-induced *hcp* formation, reaching ~11–20 % phase fraction depending on the measurement technique (**Table 1**). In contrast, the aged specimen deformed at the same temperature shows significantly less transformation (6 %), despite the lower testing temperature promoting the formation of *hcp* martensite. These results provide strong evidence that the effect of CSRO persists even when the thermodynamic driving force for TRIP is enhanced.

The corresponding (100) IPF maps further reveal extensive orientation gradients and banded microstructures across all specimens, indicating strong strain localization during deformation. However, the recrystallized condition presents more frequent transformation-related orientation variants, while the aged condition shows fewer *hcp* variants and a stronger *fcc* crystallographic continuity. This behavior suggests that aging increases the mechanical stability of the *fcc* lattice, hindering the dissociation of perfect dislocations and reducing the availability of Shockley partials, which are necessary for the *fcc→hcp* transformation.

**Figure 6** shows representative TEM microstructural features of the recrystallized and aged conditions after deformation at 173 K. The high-angle annular dark-field (HAADF) image of a recrystallized sample (**Fig. 6a**) reveals a deformation microstructure dominated by dense, crystallographically aligned planar defects forming long, parallel lamellae extending over several tens to hundreds of nanometers. Such features are characteristic of stacking faults and deformation twins commonly activated in low SFE *fcc* alloys [39]. The surrounding contrast variations suggest a high local dislocation density and significant lattice distortion associated with intense planar slip.

The corresponding selected area electron diffraction (SAED) pattern (**Fig. 6b**) is indexed to an *fcc* lattice along the $[110]_{fcc}$ zone axis, based on the characteristic rectangular arrangement of the $\{111\}_{fcc}$, $\{200\}_{fcc}$, and $\{220\}_{fcc}$ reflections. Other reflections mirrored across the $\{111\}_{fcc}$ planes arise from Σ3 deformation twins, consistent with the $\{111\}\langle 112\rangle$ twinning mode. Weak extra reflections accompanied by pronounced streaking along $\langle 111 \rangle_{fcc}$ directions show a high density of stacking faults and the formation of *hcp* martensite. These reflections can be indexed assuming an *hcp* structure viewed along the $[1\bar{1}20]_{hcp}$ zone axis, obeying the orientation relationship $(111)_{fcc} \| (0001)_{hcp}$ and $[\bar{1}10]_{fcc} \| [1\bar{1}20]_{hcp}$, characteristic of a fault-mediated *fcc* → *hcp* transformation.

High-resolution TEM (HRTEM) (**Fig. 6c**) confirms the coherent nature of the planar defects, showing abrupt stacking-sequence changes across twin and stacking-fault boundaries and nanoscale regions with *hcp* stacking embedded within the *fcc* matrix. Together, these observations demonstrate consistently that plastic deformation is accommodated by partial-dislocation activity, leading to extensive stacking faulting, deformation twinning, and localized *fcc* → *hcp* transformation, consistent with TRIP/TWIP-assisted deformation in low-SFE CoCrNi-based alloys [40].

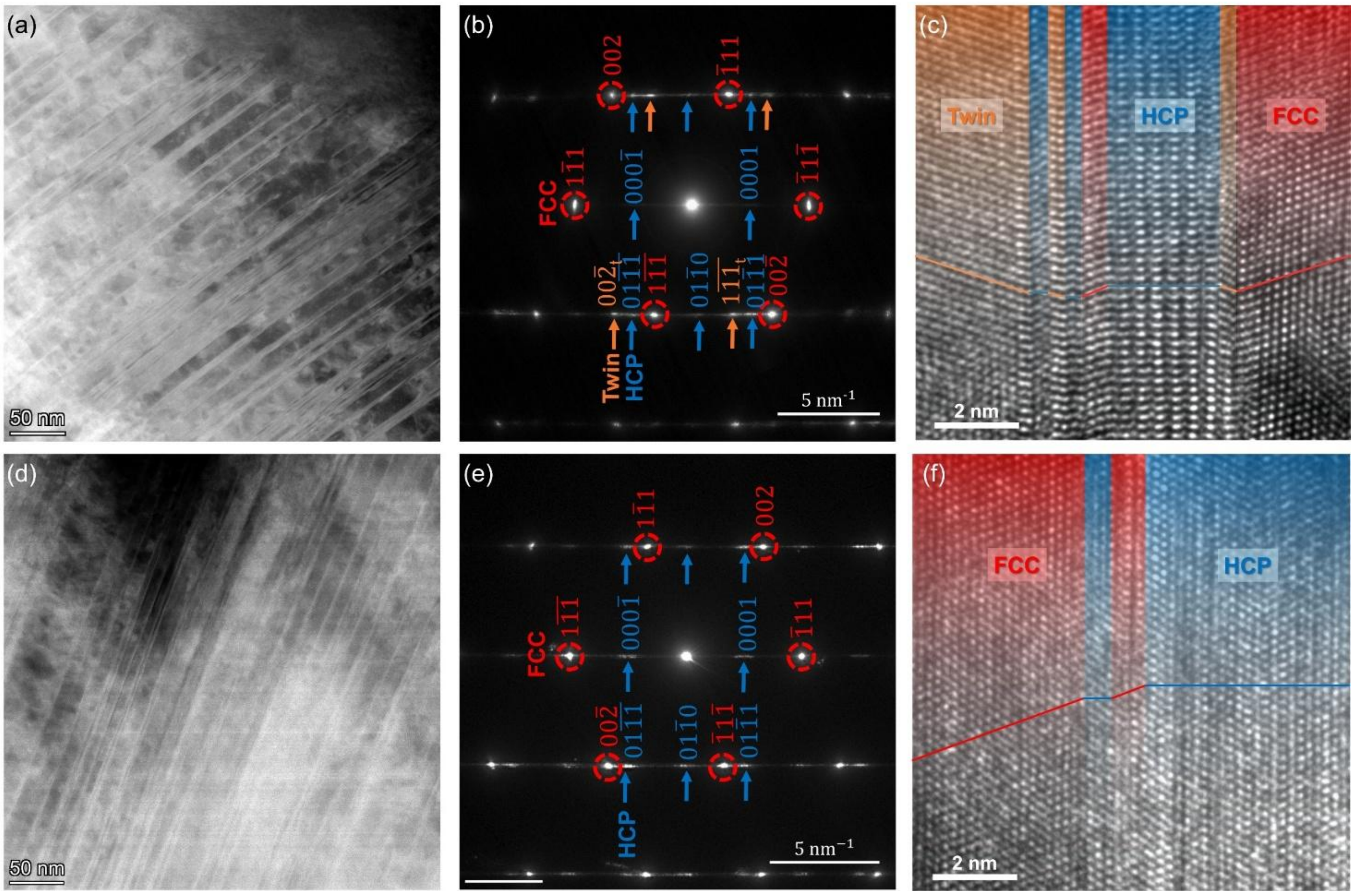

**Figure 6**. **Transmission electron microscopy characterization of deformation-induced planar defects in the *fcc* matrix following deformation at 173 K**. (**a–c**) post-deformation of recrystallized condition; (**a**) HAADF-STEM image showing a dense network of planar defects, (**b**) corresponding SAED pattern revealing *fcc* reflections with additional streaking and extra spots associated with faulting, twinning, and *hcp* lamellae, and (**c**) HRTEM image confirming nanoscale stacking faults, twin boundaries, and *fcc* → *hcp* transformation products; (**d–f**) post-deformation of aged condition: (**d**) HAADF-STEM image, (**e**) SAED pattern, and (**f**) HRTEM image.

A similar description applies to **Figs. 6d–f**, which reveal qualitatively the same types of deformation-induced planar defects in the aged condition evaluated at 173 K. HAADF-STEM, SAED, and HRTEM observations confirm the presence of stacking faults and thin *hcp* lamellae within the *fcc* matrix. However, the SAED image (**Fig. 6e**) does not reveal the presence of twin mirrored reflections. Nonetheless, due to the inherently small, probed volume in TEM, it is not possible to draw statistically meaningful conclusions on potential differences in lamella thickness or the overall volume fraction of the *hcp* phase and twinnability between the recrystallized and aged conditions. Although the lamella in the aged condition appears qualitatively finer (**Fig. 6d**), this observation should be regarded as suggestive rather than conclusive and it is not used to quantitatively assess differences in transformation extent in this study.

**Discussion**

The present experimental results indicate that CSRO modifies the intrinsic planar fault energetics of the $Co_{30}Cr_{40}Ni_{30}$ alloy and thereby alters its deformation pathways (**Figs. 3-6**). Because the recrystallized and aged conditions exhibit comparable grain-size distributions, crystallographic textures, and phase constitutions (**Fig. 1**), the observed differences in TRIP activity are therefore unlikely to originate from major differences in these microstructural variables and instead point to a role of atomic-scale chemical correlations. DSC measurements (**Fig. 2**) confirm that aging at 747 K produces a significant degree of CSRO, consistent with previous thermophysical studies of CoCrNi-based alloys [10,22]. The exothermic and endothermic signatures observed during heating correspond to reversible ordering and disordering processes, demonstrating that the aged condition contains a thermodynamically stabilized short-range ordered structure.

**Figure 7a** compares representative atomic configurations of the random and chemically equilibrated $Co_{30}Cr_{40}Ni_{30}$ alloy after MC simulation. Despite the development of CSRO, both configurations remain visually indistinguishable, reflecting the statistical nature of CSRO [41]. Consequently, its presence requires a quantitative description through the WC parameters.

The evolution of first-nearest-neighbor WC parameters $\alpha_{\mathrm{ij}}$ during the MC simulations (**Fig. 7b**) reveals the progressive development of CSRO from an initially random solid solution toward a stable equilibrium configuration. The WC parameters reach stationary distribution after approximately 30,000 MC steps. The equilibrium of the WC parameters (**Fig. 7c**) shows distinct pair-specific chemical preferences, with negative values indicating a preference for unlike atomic neighbors and positive values reflecting the avoidance of such pairs. These correlations are consistent with previously reported ordering tendencies in CoCrNi-based alloys [15,18], showing energetically favorable Co–Cr and Ni–Ni interactions, while Cr-Cr and Co-Ni pairs avoidance, which reduce the alloy enthalpy while decreasing configurational randomness [17,22]. Convergence metrics of the hybrid MC/MD simulation at 750 K are shown in **Fig. S8** in the Supplementary Information.

The development of CSRO substantially modifies the generalized planar-fault-energy (GPFE) surface (**Fig. 7d**). Relative to the random configuration, the CSRO configuration exhibits higher stable stacking-fault ($\gamma_{sf}$), unstable stacking-fault ($\gamma_{us}$), and unstable twinning-fault ($\gamma_{ut}$) energies. The increase in $\gamma_{sf}$ reduces the equilibrium separation of Shockley partials, while the higher $\gamma_{us}$ increases the energetic barrier for partial-dislocation nucleation. Together, these changes make the formation and accumulation of extended stacking faults less energetically favorable, providing an atomistic basis for the reduced propensity for the sequential partial-dislocation activity associated with *fcc*→*hcp* transformation.

This atomistic picture provides a consistent explanation for experimental observations. At both 300 and 173 K, the CSRO-enriched condition develops substantially less deformation-induced *hcp* martensite than the CSRO-lean condition (**Figs. 4 and 5**). In CoCrNi-based alloys, the *fcc*→*hcp* transformation is associated with the accumulation of stacking faults generated by successive Shockley partials. The higher planar-fault energies predicted for the CSRO configuration therefore provide a mechanistic basis for suppressing this transformation. This effect is particularly significant at 173 K, where lowering the temperature normally promotes deformation-induced *hcp* transformation through reduced SFE and an increased thermodynamic driving force for transformation [3]. The persistence of a substantially lower *hcp* fraction in the CSRO-enriched condition demonstrates that CSRO can counteract this temperature-induced shift toward transformation.

The mechanical response further supports this interpretation. At 300 K, the two conditions exhibit broadly similar tensile behavior despite their different transformation fractions, indicating that the effect of CSRO on the macroscopic response is relatively subtle under these conditions. At 173 K, where deformation-induced transformation contributes more strongly to strain hardening, the higher *hcp* fraction in the CSRO-lean condition is accompanied by greater strain hardening and ductility, whereas suppression of TRIP in the CSRO-enriched condition reduces its hardening capacity. Thus, CSRO influences the mechanical response primarily by altering the relative contribution of TRIP rather than by substantially changing the initial strength.

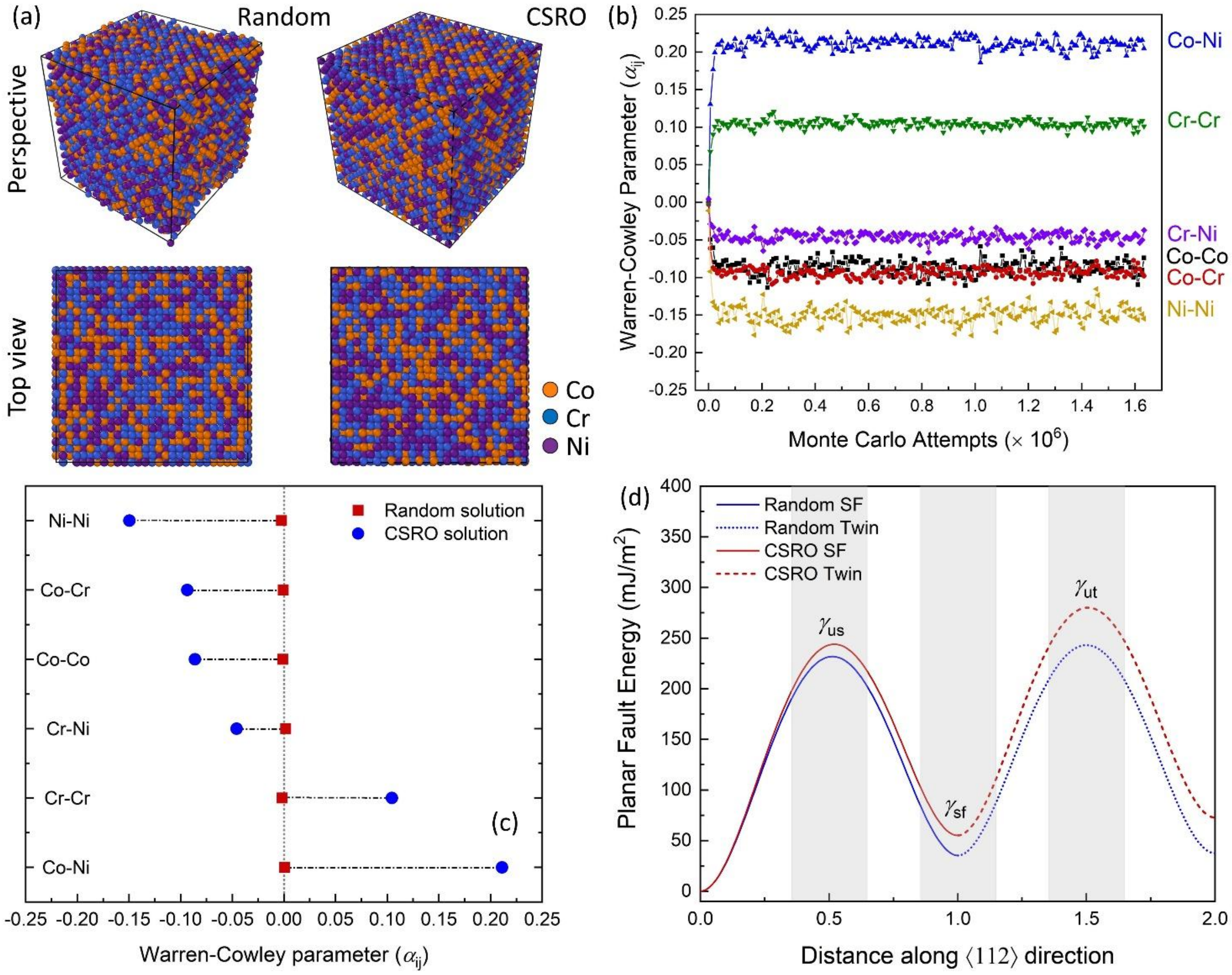

**Figure 7. Hybrid Monte Carlo/Molecular Dynamics simulation results and generalized planar fault energy calculations for the random and CSRO $Co_{30}Cr_{40}Ni_{30}$ alloy**. **(a)** Representative atomic configurations of the random and equilibrated CSRO supercells after MC simulations; **(b)** evolution of the first-nearest-neighbor WC parameters as a function of MC total attempts, showing the development and convergence of CSRO; **(c)** equilibrium WC parameters for the first-nearest-neighbor atomic pairs, comparing the random and CSRO configurations. Negative and positive WC values indicate preferred and avoided atomic pair correlations, respectively; **(d)** GPFE curves calculated at 0 K for the random and CSRO configurations, illustrating the effect of CSRO on the fault energetics.

Picak *et al*. [27] reported that CSRO promotes the concurrent activation of TWIP and TRIP during tensile deformation of equiatomic CoCrNi single crystals along specific crystallographic orientations, apparently contrasting with the present findings. This difference may arise from the distinct thermal histories used to establish the chemical-ordering state. Here, CSRO was developed by aging at 748 K for 240 h followed by water quenching, whereas Picak *et al*. [27] employed annealing at 1273 K followed by furnace cooling. Because CSRO depends on both temperature and the kinetics of chemical redistribution during subsequent cooling, these treatments can produce different degrees and local configurations of CSRO [8]. Thus, the high annealing temperature used by Picak et al. does not necessarily imply a higher or lower final CSRO state than that obtained here. The apparently different deformation responses may therefore reflect differences in the resulting CSRO equilibrium, in addition to differences in crystal orientation and sample geometry.

Overall, experimental and atomistic results establish a mechanistic link between CSRO, planar-fault energetics, and deformation-pathway selection. CSRO modifies the GPFE surface, increasing the energetic cost of both stable and unstable fault formation and thereby reducing the propensity for the sequential partial-dislocation activity associated with *fcc→hcp* transformation. Notably, this effect is pronounced in the transformation behavior while remaining relatively subtle in conventional tensile metrics, demonstrating that local chemical order can alter the active deformation pathway without requiring a substantial change in alloy composition or initial strength. CSRO therefore represents an additional materials-design variable for controlling deformation mechanisms. More broadly, the results suggest that manipulation of local chemical order may provide a route to tailoring transformation behavior and strain-hardening response in alloys where multiple deformation mechanisms compete.

## Methods

### *Sample processing*

The $Co_{30}Cr_{40}Ni_{30}$ (at. %) composition was selected because it has a lower SFE (15.9 mJ $m^{-2}$) than the widely investigated equiatomic CoCrNi alloy (19.3 mJ $m^{-2}$) [42]. Thus, it is more susceptible to TWIP and TRIP mechanisms [37], especially at room temperature. The alloy ingot was industrially produced by Villares Metals S.A. in a vacuum induction furnace and supplied to the Department of Materials Engineering (DEMa, UFSCar) for research purposes [37]. From the as-cast ingot, two plates with approximate dimensions of 160 × 160 × 30 mm were sectioned using a band saw. The material was then homogenized at 1373 K for 6 h in a resistance furnace under ambient atmosphere. Subsequently after homogenization, the plates were hot rolled at 1373 K, with intermittent reheating for 10 min to keep a uniform temperature throughout the process. The thickness was reduced by 60 %, from 30 to 12 mm.

Thereafter, the plates underwent an annealing heat treatment, for complete recrystallization, at 1373 K for 1 h, followed by water quenching. A subset of the specimens was then aged at 747 K for 240 h to induce significant levels of CSRO, following the protocol established by Bacurau *et al.* [22] for CoCrNi-based alloys.

***Microstructural and structural characterization***

*Metallographic preparation*

For metallographic analysis, samples were hot mounted in bakelite, ground using silicon carbide paper (120–1500 grit), and polished with 3 and 1 µm diamond suspensions and finally polished in a vibratory polishing machine with 0.05 µm colloidal silica. After each polishing step, the samples were ultrasonically cleaned in 96.7 % alcohol. The samples analyzed by EBSD, after the grinding and polishing steps, were electrochemically polished in a perchloric acid ($HClO_4$) solution (80 % absolute ethanol) for 15 seconds under an applied voltage of 30 V.

*Scanning electron microscopy*

The specimens were analyzed by scanning electron microscopy (SEM) using a high-resolution TESCAN MIRA, including energy-dispersive X-ray spectroscopy (EDS) and EBSD, to assess the microstructure, determine phase constitution, and quantify chemical composition. EBSD maps were acquired with a 70 ° sample tilt, ~16 mm working distance, 25 kV accelerating voltage, and a detector-to-sample distance of 16 mm. The EBSD data was processed and analyzed using the ATEX software [43].

*X-ray diffraction*

In-house XRD measurements were conducted in transmission mode using a molybdenum X-ray source (Anton Paar XRDynamic 500). Scans were acquired over a $2\theta$ range of 5–50° with a step size of 0.02°. For deformed samples, a $2\theta$ range of 14–28 ° (wave vector $Q \sim 2.3$–$4.3$ Å$^{-1}$) was used for increased resolution. The specimens used in transmission mode had a thickness of 100–130 µm.

Synchrotron XRD (SXRD) measurements were performed at the Paineira beamline of the Sirius synchrotron facility, Brazilian Center for Research in Energy and Materials (CNPEM). Ex situ deformed samples were analyzed under identical experimental conditions. Data were collected in the $2\theta$ range from 1–110°, with an angular resolution of 0.0025° and an integration time of 0.3 s per step using a PIMEGA 450D detector. The incident X-ray energy was 29.0814 keV (wavelength $\lambda = 0.4263$ Å). These experimental conditions enabled high-resolution diffraction data collection suitable for detailed structural analysis.

*Determination of phase volume fractions*

Phase volume fraction was quantified from the XRD data using Equation 1, where $A$ is the experimental integrated intensity for a given *hkl* reflection (i.e., the area under the diffraction peak), $I$ is the theoretical relative integrated intensity for the same *hkl* reflection [calculated using Equation (2), and $n$ is the number of peaks considered [4]. Two *fcc* reflections (111 and 200) and two or three *hcp* peaks ($10\bar{1}0$, 0002, and $10\bar{1}1$) were used in the analysis.

The theoretical relative integrated intensity ($I$) was calculated using Equation 2, where $V$ is the unit cell volume ($m^3$), $|F|$ is the structure factor, $p$ is the multiplicity factor, and the term in parentheses corresponds to the Lorentz–polarization factor, dependent on diffraction angle ($\theta$) [4].

$$f_{\mathrm{hcp}} = \frac{\frac{1}{n}\sum_{j=1}^{n}\left(\frac{A_{\mathrm{hcp}}^{j}}{I_{\mathrm{hcp}}^{j}}\right)}{\frac{1}{n}\sum_{j=1}^{n}\left(\frac{A_{\mathrm{hcp}}^{j}}{I_{\mathrm{hcp}}^{j}}\right)+\frac{1}{n}\sum_{j=1}^{n}\left(\frac{A_{\mathrm{fcc}}^{j}}{I_{\mathrm{fcc}}^{j}}\right)} \quad \text{(Eq. 1)}$$

$$I = \frac{1}{V^2}\,|F|^2\; p\left(\frac{1+\cos^2 2\theta}{\sin^2\theta\cos\theta}\right) \quad \text{(Eq. 2)}$$

*Transmission electron microscopy*

Electron-transparent specimens were prepared from mechanically sectioned samples that were initially ground and polished to a thickness of approximately 60 μm. Final thinning to electron transparency was achieved by ion milling. TEM imaging was conducted using two field-emission instruments run at an accelerating voltage of 200 kV: a Thermo Fisher Scientific Talos F200X equipped with a cold field-emission gun and a Talos F200i. Both instruments were used for high-angle annular dark-field scanning TEM (HAADF-STEM) imaging, selected-area electron diffraction (SAED), and high-resolution scanning TEM (HRSTEM) analysis.

***Thermal analysis***

*Differential scanning calorimetry*

To indirectly assess the formation of CSRO in the alloy, variations in specific heat capacity $c_{\mathrm{p}}$ were evaluated through DSC measurements [10,22]. Samples in the recrystallized and aged conditions were analyzed to investigate the formation and dissolution of CSRO across two heating cycles [10]. Sapphire was used as reference material, and measurements were performed using platinum crucibles for increased sensitivity. The $c_{\mathrm{p}}$ curves were obtained for two specimens under each condition, considering both heating cycles. The samples weighed approximately 40 mg.

During both heating cycles, the samples were heated from 298 K to 973 K at a rate of 20 $K\cdot min^{-1}$. Isothermal holds of 10 min were applied at 373 K and 973 K to ensure thermal equilibration and a uniform temperature distribution throughout the sample. The $c_{\mathrm{p}}$ behavior was analyzed exclusively during the heating segments, while the cooling segments were excluded from the analysis.

*Enthalpy estimation*

To estimate the enthalpy associated with the formation or dissolution of CSRO in the alloy, the calorimetric heat $Q_{\mathrm{total}}$ evolution was evaluated using Equation 3 [22], where $\Delta H_{\Delta T}$ represents the enthalpy change solely from temperature variation, obtained by integrating the heat capacity over the selected temperature interval, and $\Delta H_{\mathrm{CSRO}}$ corresponds to the enthalpy change associated with the formation or dissolution of CSRO [22]. The anomalous deviations in heat-capacity behavior are attributed to significant enthalpy contributions from short-range ordering effects [10], since no phase transformations or other thermally activated reactions are expected to occur within this temperature range for the chosen alloy.

$$Q_{\mathrm{total}} = \int_{T_i}^{T_f} q \, dT = \Delta H_{\Delta T} + \Delta H_{\mathrm{CSRO}} \qquad \text{(eq. 3)}$$

From the curves obtained for the recrystallized condition, the enthalpy associated with CSRO formation could be quantified from the observed exothermic peaks, while CSRO dissolution in aged samples could be quantified from the observed endothermic peaks.

**Mechanical properties testing**

Tensile specimens were machined by electrical discharge machining into flat dog-bone subsize sheet-type geometry with a thickness of 1.0 mm. The loading axis was aligned parallel to the rolling direction. The specimens' dimensions are provided in **Figure S1** of the Supplementary Information. Each condition was assessed in triplicate to ensure reproducibility.

Uniaxial quasi-static tensile tests were conducted on an Instron 5500R universal testing machine at a strain rate of $10^{-3}$ $s^{-1}$. Tests were performed at room temperature (300 K) and at a cryogenic temperature (173 K) using an Instron environmental chamber attached to the machine and cooled using liquid nitrogen.

**Atomistic simulations**

*Machine-learning interatomic potential*

Atomistic simulations were performed for an *fcc* $Co_{30}Cr_{40}Ni_{30}$ (at. %) composition using the Large-scale Atomic/Molecular Massively Parallel Simulator (LAMMPS) code [44] with a PACE/ACE machine-learning interatomic potential previously developed for the Co-Cr-Ni system [45]. The simulation cell consisted of a 14×14×14 *fcc* supercell containing 10,976 atoms. The initial configuration was generated as a chemically random solid solution.

*Thermal equilibration*

The chemically random solid solution (RSS) was first equilibrated in the isothermal-isobaric (NPT) ensemble at 750 K and zero hydrostatic pressure. A timestep of 1 fs was employed for 50 ps. Temperature and pressure were controlled using a Nosé-Hoover thermostat and barostat with damping parameters of 0.1 ps for both variables. Ten independent runs were used for WC statistics.

The first-nearest-neighbor WC parameter $\alpha_{ij}$ was computed from the probability $p_{ij}$ of finding an atom $j$ in the first-nearest-neighbor shell of an atom $i$, normalized by the global concentration $c_j$ of element $j$:

$$\alpha_{ij} = 1 - {p_{ij}}/{c_j} \qquad \text{(eq. 4)}$$

Thermal-equilibrium chemical configurations were subsequently obtained using hybrid Monte Carlo/ molecular dynamics (MC/MD) simulations at 750 K. For this, the fixed atom/swap command was used with trial swaps between atoms of different chemical types. The swap acceptance was determined by the Metropolis-Hastings algorithm [46] contained within LAMMPS. Each hybrid cycle consisted of 150 attempted swaps followed by 50 molecular dynamics steps in the NPT ensemble using the same 1 fs timestep. The hybrid MC/MD simulation was carried out until 50 accepted sweeps were reached, corresponding to each atom participating in 100 accepted swaps on average.

For both the RSS and equilibrated (CSRO) systems the overall properties and WC were only averaged during steady-state. Simulation cells and atomic configurations were visualized using the OVITO software after fixed-cell relaxation by 10 energy-minimization iterations [44].

*Generalized planar fault energy*

To evaluate the influence of CSRO on deformation mechanisms, generalized planar fault energy curves were calculated for both the random and CSRO-enriched microstructures at 0 K by rigidly displacing the upper half of the simulation cell across the (111) plane along the ⟨112⟩ direction.

## Acknowledgements

A.F.A. acknowledges the financial support from the São Paulo Research Foundation (FAPESP), grants 2024/11008-7 and 2024/17093-6, the National Council for Scientific and Technological Development (CNPq), grants 157806/2025-1 and 153727/2024-1, and the Wallenberg Initiative Materials Science for Sustainability (WISE). G.C.S. was supported by FAPESP under grants 2023/07403-5 and 2025/08513-4. G.B was supported by FAPESP under grants 2021-05408-4 and 2025-23860-2. V.P.B. was supported by FAPESP under grant 2024/02515-2. EMM acknowledges CNPQ, grant 403997/2025-9. F.G.C. acknowledges FAPESP, grants 2024/01818-1 and 2022/02770-7, and CNPq, grants 403832/2023-3 and 444393/2024-2. To CAPES – Brazilian Federal Agency for Support and Evaluation of Graduate Education – for their financial support in this work through a scholarship, process no. 88887.952714/2024-00. The research was sponsored by the Army Research Office and was accomplished under Grant Number W911NF-23–1–0310. The views and conclusions contained in this document are those of the authors and should not be interpreted as representing the official policies, either expressed or implied, of the Army Research Office or the U.S. Government. The U.S. Government is authorized to reproduce and distribute reprints for Government purposes notwithstanding any copyright notation herein. This research used facilities of the Brazilian Synchrotron Light Laboratory (LNLS), part of CNPEM, a private non-profit organization under the supervision of the Brazilian Ministry for Science, Technology, and Innovations (MCTI). The Paineira beamline staff is acknowledged for their aid during beamtime 20252573.